\documentclass[
    ,final            
  ]
  {aipproc}

\layoutstyle{6x9}

\newcommand{\be}{\begin{equation}}
\newcommand{\ee}{\end{equation}}
\newcommand{\ba}{\begin{array}{c}}
\newcommand{\ea}{\end{array}}
\newcommand{\bqa}{\begin{eqnarray}}
\newcommand{\eqa}{\end{eqnarray}}
\begin{document}

\title{Is the $\sigma$ meson dynamically generated?\footnote{Talk presented by  Zheng at ``Quark Confinement
      and Hadron Spectrum VII'', 2--7 Sept. 2006,  Ponta Delgada, Acores, Portugal}}



\author{Zhi-hui Guo, L.~Y.~Xiao and  H.~Q.~Zheng}{
  address={Department of Physics, Peking University, Beijing 100871, P.~R.~China}
}

\begin{abstract}
 We study the problem whether the $\sigma$ meson is
 generated `dynamically'. A pedagogical analysis on the toy O(N) linear sigma model
 is performed and we find that the large $N_c$ limit and the
 $m_\sigma\to \infty$ limit does not commute.
 The sigma meson may not
 necessarily be described as a dynamically generated resonance. On the
 contrary, the sigma meson may be more appropriately  described by
 considering it as an explicit degree of freedom in the effective lagrangian.
\end{abstract}

\maketitle

The long standing debate on whether there exists a light and broad
resonance (for the historical reason named as the `$\sigma$'
meson~\cite{Tornqvist}) has finally come to an end. The Roy equation
analysis clearly indicates a light and broad resonance pole located
just inside the analyticity domain established from Roy
equations~\cite{CCL06}. Despite that there may still exist some
disputes at technical level on pole locations, Roy equation analysis
starts from first principles of quantum field theory and the result
on the existence of the $\sigma$ pole is robust. The very existence
of the sigma pole can actually be understood in a simper,  more
intuitive and also rigorous way: one writes a dispersion relation
for the analytic continuation of $\sin(2\delta_\pi)$ in the I,J=0,0
channel of elastic $\pi\pi$ scattering and use chiral perturbation
theory to estimate the background (left hand cut) contribution. In
this way one finds that the background contribution to the phase
shift ($\sin(2\delta_\pi)$) is negative and concave whereas the
experimental data is positive and convex~\cite{xz00}. The difference
can only be made up by a pole contribution, according to the
standard $S$ matrix theory principle. On the other side, the
$\sigma$ pole location found from different dispersive
analyses~\cite{CCL06,zhou05} agree with each other, within error
bars, hence proving convincingly the stability of the numerical
outputs on the pole location from dispersive analyses. Also the
results from dispersive analyses are also in fair agreement with
recent experimental determinations~\cite{sigmaexp}. On the other
side, dispersive analyses also reveal, in a model independent way,
the existence of a light and broad resonance named
$\kappa$~\cite{zhou06,moussallam}, which are again in agreement with
the recent experimental determinations~\cite{expkappa}. In the
partial wave dispersion relation for $\pi K$ scattering, there
exists the circular cut due to un-equal mass kinematics. In
Ref.~\cite{zhou06}, the background contribution is estimated along
the outer edge of the circular cut, hence the complicated cut
structure inside the circular cut is avoided.

Having firmly established the existence of the light and broad
$\sigma$ and $\kappa$, the next important question is what are these
resonances? There exist different proposals to explain these
resonances. For example, there are tetra quark model~\cite{tetra},
linear sigma model at hadron level~\cite{LsigmaH}, linear sigma
model at quark level~\cite{LsigmaQ}, and also the ENJL
model~\cite{ENJL}. Also there exists the approach to explain the
light scalars as dynamically generated resonances~\cite{pelaez}. In
Ref.~\cite{xiaoly06} it is observed that since the widths of
$\sigma$ and $\kappa$ are very large the pole mass and and `line
shape' mass are very different quantities, one should be extremely
careful when discussing the mass relations between scalars: the tree
level mass relations have to be among those line shape masses rather
than those pole masses. The mass relations obtained from the
extended Nambu Jona-Lasinio lagrangian are discussed and it is
argued that the lightest scalars, $\sigma$, $\kappa$, $a_0(980)$ and
$f_0(980)$ form an nonet, as the chiral partners of the
pseudo-goldstone bosons~\cite{xiaoly06}. The mass relations are
crude, but it is expected that they grasp the major characters of
the physics underlined. The major difficulty in this approach is to
explain the large  mass of $f_0(980)$~\cite{xiaoly06}, and more
detailed dynamical analysis may be needed~\cite{baru}.

In the approach to consider the lightest scalars as chiral partners
of the pesudo-goldstone bosons, one thing remains to be explained is
how to understand the approach that the $\sigma$ and $\kappa$ are
generated dynamically. Owing to the complexity of the problem, we in
the following discuss the unitarization approximations to the
solvable O(N) $\sigma$ model. As will be shown later, it will be
helpful to understand several difficult issues.

The O(N) linear $\sigma$ model lagrangian is
 \be\label{lag1}
 \mathcal{L}=\frac{1}{2}\partial_\mu\Phi^T\partial^\mu\Phi-\frac{1}{2}m^2\Phi^T\Phi-\frac{\lambda}{8N}(\Phi^T\Phi)^2
\ee where $\Phi=(\Phi_1,\Phi_2,\cdots,\Phi_N)^T$. The explicit
symmetry breaking interaction is characterized by,
  \be\label{Lsb}
\mathcal{L}_{S.B.}=v\,m_\pi^2\Phi_N\ .
   \ee
Here we treat the lagrangian as a cutoff effective lagrangian. That
is, in our calculation we make the following replacement:
 \be
\Gamma(\epsilon)+\ln{4\pi}+\ln{\frac{\mu^2}{m_\pi^2}}\Rightarrow
\ln{\frac{\Lambda^2}{m_\pi^2}}\ .
 \ee
 It has been proved that in such a toy model the [n,n] Pad\'e
 amplitudes reproduce the
 exact sigma pole location and the K matrix unitarizations are good
approximations~\cite{Willenbroch1990}.  Nevertheless, such a nice
property is not maintained if the pion fields are expressed in the
non-linear representation, since for the latter the chiral expansion
series has to be truncated. There are variants of Eq.~(\ref{lag1}).
For example one may make a polar decomposition to the linear
lagrangian and recast it into the following form:
 \be\label{polar}
\mathcal{L}_{polar}=\mathcal{L}^{\sigma}+\frac{1}{2}(1+\frac{\sigma}{v})^2(\partial_\mu\vec{\pi}\cdot\partial^\mu\vec{\pi}
+\partial_\mu\sqrt{v^2-\vec{\pi}\cdot\vec{\pi}}\partial^\mu\sqrt{v^2-\vec{\pi}\cdot\vec{\pi}})
\ee where
  \be
\mathcal{L}^{\sigma}=\frac{1}{2}\partial_\mu\sigma\partial^\mu\sigma-\frac{1}{2}m^2(\sigma+v)^2-\frac{\lambda}{8N}(\sigma+v)^4\
. \ee One further expands the square root in Eq.~(\ref{polar}) when
calculating scattering amplitudes. Also one may completely neglect
the sigma field in   Eq.~(\ref{polar}) to get the non-linear sigma
model,
 \be\label{mod}
  \mathcal{L}_{
NL}=\frac{1}{2}(\partial_\mu\vec{\pi}\cdot\partial^\mu\vec{\pi}
+\partial_\mu\sqrt{v^2-\vec{\pi}\cdot\vec{\pi}}\partial^\mu\sqrt{v^2-\vec{\pi}\cdot\vec{\pi}})\
. \ee
 Or one integrate out the sigma field at tree level to get the
modified non-linear sigma model lagrangian,
 \be \label{mod'}
\mathcal{L}_{\overline{NL}}=\mathcal{L}_{NL}+\frac{1}{2m_\sigma^2v^2}[(\partial_\mu\vec{\pi}\cdot\partial^\mu\vec{\pi})^2
-m_\pi^2\vec{\pi}\cdot\vec{\pi}\partial_\mu\vec{\pi}\cdot\partial^\mu\vec{\pi}+\frac{m_\pi^4}{4}(\vec{\pi}\cdot\vec{\pi})^2]
.\ee
 We have tested various unitarization approximations and the
details will be given elsewhere. Here we only briefly discuss
 the properties of  [1,1] Pad\'e  amplitudes constructed using
${\cal L}_{polar}$, ${\cal L}_{NL}$, ${\cal L}_{\overline{NL}}$,
respectively. Notice that here we work in the cutoff version of
effective lagrangian and hence no counter term is needed when one
make calculations at 1-loop level. In each amplitude, a pole is
found close to or not far from the sigma pole of the original
lagrangian Eq.~(\ref{lag1}). In the case of ${\cal L}_{polar}$, the
pole found in the unitarized amplitude is not dynamical. For ${\cal
L}_{NL}$, ${\cal L}_{\overline{NL}}$, the poles are called
`dynamical'. Except these `$\sigma$' poles being reproduced, there
may exist other spurious poles. The spurious pole does not occur in
the lagrangian with linearly realized chiral symmetry, hence one may
find that the lagrangian with linearly realized chiral symmetry are
better for the purpose of unitarization. Another lesson one may
learn is that a `dynamically generated' resonance may or may not be
truly dynamical. For ${\cal L}_{NL}$, the `$\sigma$' pole is indeed
dynamical, but for ${\cal L}_{\overline{NL}}$ the `$\sigma$' pole
just regenerates the $\sigma$ particle being integrated out in the
original lagrangian Eq.~(\ref{lag1}). For the latter case, the
$\sigma$ is, of course, better (or more conveniently) described
 by explicitly including it in the lagrangian.

The dynamical poles generated from Eq.~(\ref{mod}) and
Eq.~(\ref{mod'}) have quite different dynamical properties, however.
It is not difficult to check that the pole location of the the
`$\sigma$' pole produced by Eq.~(\ref{mod}) is $\sqrt{s_p}\propto
v=f_\pi\propto \sqrt{N_c}$ and moves to infinity when
$N_c\to\infty$, whereas the pole generated from Eq.~(\ref{mod'})
behaves as $\sqrt{s_p}\to m_\sigma$ when $N_c\to\infty$. Apparently
only the latter is correct when simulating Eq.~(\ref{lag1}). The
lesson one may learn from here is that the large $N_c$ limit and the
$M_\sigma\to \infty$ limit do not commute.

O(N) model is only a simple toy model, comparing with the
complicated structure of QCD. However one may still learn some
useful lessons from above. The Eq.~(\ref{mod}) simulates the current
algebra non-linear sigma model in reality whereas Eq.~(\ref{mod'})
resembles $O(p^4)$ chiral perturbation theory lagrangian in reality.
In the real situation, Actually similar things happen. The current
algebra prediction to the $\sigma$ pole location~\cite{leutwyler05}
 \be
\sqrt{ s_{\sigma}} \simeq \sqrt{16i\pi f_\pi^2}\simeq 463-463i\ ,\ee
which moves to $\infty$ when $N_c\to\infty$,
 may receive important corrections:
 \be\label{pade11}
 s_{\sigma} \simeq \frac{16i\pi f_\pi^2}{1+16i\pi f_\pi^2\triangle}\
 ,
\ee where $\triangle=\frac{2}{3f_\pi^2}(22L_1+14L_2+11L_3)\propto
O(N_c^0)$.
 The above expression is obtained from [1,1] Pad\'e
approximation in the chiral limit~\cite{sun05}. Hence it was not
clear what approximation is made in obtaining Eq.~(\ref{pade11}). It
is however also obtainable using the PKU parametrization form under
two assumptions in the large $N_c$ and chiral limit~\cite{sun05}: 1)
one pole (the `$\sigma$' pole) dominance in the $s$ channel, 2)
neglecting all resonance exchanges in the crossed channels, which
can also be at the leading order in  $1/N_c$ expansion. Even though
Eq.~(\ref{pade11}) is only a rough approximation, we expect it gives
the correct $N_c$ dependence of the sigma pole, if the `$\sigma$'
meson contributes to the low energy constants when $N_c$ is
large~\cite{xiao05}. For more detailed discussion related to the
$\sigma$ pole location in the large $N_c$ limit one is referred to
Ref.~\cite{sun05}.

Since the `$\sigma$' meson found in the [1,1] Pad\'e approximation
finally falls down to the real axis in the large $N_c$ limit, it is
suggested, through the analysis given above, that the `$\sigma$'
meson is a true particle, i.e., the $\sigma$ meson being responsible
for the spontaneous chiral symmetry breaking in the linear
realization of chiral symmetry.

 {\bf Acknowledgement:}  This work supported in part by China National
Natural Science Foundation
under grant number 10575002 and 
 10421503.

\end{document}